\documentclass[a4paper,11pt]{article}

\usepackage{graphicx}
\usepackage{subfigure}
\usepackage{dcolumn}
\usepackage{bm}
\usepackage{epsfig}
\usepackage{epsf}
\usepackage{amsmath,amsfonts,amssymb}

\newcounter{definition}

\def\thetable{\arabic{section}.\arabic{table}}
\def\thefigure{\arabic{section}.\arabic{figure}}
\def\theequation{\thesection.\arabic{equation}}
\def\appendix{
  \setcounter{section}{0}
  \setcounter{subsection}{0}
  \par
  \def\thesection{Appendix \Alph{section}}
  \def\theequation{\Alph{section}.\arabic{equation}}
  \def\thefigure{\Alph{section}.\arabic{figure}}
  \def\thetable{\Alph{section}.\arabic{table}}
}
\newcommand{\nc}{\newcommand}
\nc{\rnc}{\renewcommand}
\nc{\be}{\begin{eqnarray}}
\nc{\ee}{\end{eqnarray}}
\def\refeq#1{(\ref{#1})}
\def\d{\mbox d}

\def\nn{\nonumber}

\def\ip{\int_{0}^{\infty}}

\def\G{\Gamma}
\def\D{\Delta}

\def\L{\Lambda}
\def\la{\lambda}
\def\Or{\mathcal O}
\def\g{\gamma}
\def\t{\theta}
\def\al{\alpha}

\def\k{\kappa}
\def\ve{\varepsilon}
\def\vp{\varphi}
\def\vt{\vartheta}
\def\l{\left}
\def\r{\right}
\def\te{\mbox{e}}
\def\rmi{{\rm i}}

\def\bra{\langle}
\def\ket{\rangle}
\nc{\ds}{\displaystyle}
\nc{\ch}{\cosh}
\nc{\sh}{\sinh}
\nc{\vac}{|0\ket}
\nc{\zcor}[2]{\bra S_{#1}^z S_{#2}^z \ket}
\nc{\zzcor}[1]{2^#1 \, \bra \prod_{j=1}^#1 S_j^z \ket}
\rnc{\(}{\left(}
\rnc{\)}{\right)}
\rnc{\[}{\left[}
\rnc{\]}{\right]}
\title{
String correlation functions of the spin-1/2 Heisenberg XXZ chain} 
\author{
  Michael Bortz ${}^1$\thanks{michael.bortz@anu.edu.au}, \ \
  Jun Sato ${}^2$\thanks{junji@issp.u-tokyo.ac.jp}, \ \ 
  Masahiro Shiroishi${}^2$\thanks{siroisi@issp.u-tokyo.ac.jp}
\\
 \\ \it
   ${}^1$Department of Theoretical Physics, Research School of Physics and Engineering,\\ 
   Australian National University, Canberra ACT 0200, Australia 
   \\\it
  ${}^2$Institute for Solid State Physics, University of Tokyo,\\\it 
  Kashiwanoha 5-1-5, Kashiwa, Chiba 277-8581, Japan\\\it
}

\begin{document}
\maketitle
\begin{abstract}
We calculate certain string correlation functions, originally introduced
as order parameters in integer spin chains, for the spin-1/2 XXZ Heisenberg
chain at zero temperature and in the thermodynamic limit. For small distances,
we obtain exact results from Bethe Ansatz and exact diagonalization, whereas in 
the large-distance limit, field-theoretical
arguments yield an asymptotic algebraic decay. We also make contact with
two-point spin-correlation functions in the asymptotic limit.
\end{abstract}

\section{Introduction}
Haldane's work \cite{Haldane83a,Haldane83b} on the different ground-state properties 
of integer-$S$ and half-integer-$S$ spin chains triggered efforts to seek for a quantitative 
understanding of the gapped ground state of integer-$S$ chains. Among these are the works of 
den Nijs and Rommelse \cite{Nijs89} as well as Oshikawa \cite{Oshikawa92}, where the following 
generalized string correlation function was considered:
\be
\Or(n,\t)&\equiv&-4\l \langle S_1^z \exp\l[\rmi
  \theta \sum_{k=2}^{n-1} S_k^z\r] S_n^z\r\rangle \label{ordef}.
\ee
The authors of \cite{Nijs89} introduced $\lim_{n\to\infty}\Or(n,\pi)$ as an
order parameter that characterizes the gapped ground state for the $S=1$
Heisenberg chain and acquires a non-zero value there. Kennedy and Tasaki
\cite{Kennedy92} introduced a transformation showing that this is due to a broken hidden $Z_2\times Z_2$ symmetry of the model. 
In \cite{Oshikawa92}, an attempt was made to generalize this argument to integer $S>1$ chains. In the same reference, the den Nijs-Rommelse order parameter was considered for $\t\neq \pi$. Several 
subsequent works considered the generalized string correlation functions (\ref{ordef}) for integer 
spin ${S>1}$ and generic $\theta$, where $\lim_{n\to\infty}\Or(n,\t)$ acquires non-zero values. The exact calculation for the Valence Bond Solid (VBS) state shows that the correlation takes its maximum values near ${\t=\pi/S}$ \cite{Oshikawa92,Totsuka95}.

Whereas in these works, the focus was mainly on integer spin chains  motivated by Haldane's conjecture, 
interest at the same time rose for $\Or(n,\pi)$ in half-integer spin chains. Hida \cite{Hida92} studied $\Or(n,\pi)$ for alternating spin-1/2 systems, as this model describes a crossover 
between the gapped $S=1$ phase and the isotropic spin-1/2 Heisenberg chain. In that paper, he reported 
the asymptotic form of $\Or(n,\pi)\sim \mbox{const. }n^{-1/4}$ close to the uniform, isotropic spin-1/2 
chain by means of a field-theoretical approach (the constant was not known there). This means that the string 
correlation function for the spin-1/2 Heisenberg chain $\Or(n,\pi)$ decays in an 
algebraic way much slower than the usual spin-spin correlation function.

Recently, a related string correlation function
\be
\rho(n,\t) &\equiv& \l\langle \exp\l[\rmi \theta \sum_{k=1}^n
    S_k^z\r]\r\rangle \label{rho}
\ee
was introduced by Lou et al.\cite{Lou03}. They came to the conclusion that asymptotically, for spin $S=3/2$, 
$\l.\Or(n,\t)\r|_{S=3/2}\sim -\sin^2(\t/2) \,
\l.\rho(n,\t)\r|_{S=1/2}$. This means that the scaling behaviour of $\l.\rho(n,\t)\r|_{S=1/2}$ is also important for $S=3/2$, which is supported by the fact that the spin-3/2
and spin-1/2 chains are considered to belong to the same universality class \cite{Affleck87,Hallberg96}. Using a field-theoretical approach, the authors of \cite{Lou03} found
$\l.\rho(n,\t)\r|_{S=1/2}\sim \mbox{const. }n^{-\t^2/(4\pi^2)}$ with an
unspecified constant, again for the isotropic spin-1/2 chain. 

As far as two-point correlation functions of the spin-1/2 chain are concerned,
enormous progress has been made in the last decade to obtain exact expressions
from Bethe Ansatz \cite{Bethe31,KorepinBook,TakahashiBook} for short distances \cite{JimboBook,Jimbo96,Kitanine00,Kitanine02,Kitanine05,Boos01,Sakai03,Kato04,Boos03,
Boos05,Sato05,Sato06,BJMST05,BJMST06n1,BJMST06n2}
and from field-theory for both the amplitudes and the exponents of the leading terms in
the asymptotic limit \cite{Affleck88,Affleck98,Lukyanov98,Lukyanov03}. These results are not
restricted to the isotropic point, but cover the critical anisotropic regime
as well,
\be
H=J \sum_{l=1}^N \( S^x_l S^x_{l+1}+S^y_l S^y_{l+1} + \D S^z_l S^z_{l+1}\) \label{XXZ_Hamiltonian},
\ee
with periodic boundary conditions and $J>0$. In the following, we use the anisotropy 
parameter $\g$ to parameterize the anisotropy $\D=:\cos\g$, with $0<\g<\pi$, such that the isotropic points $\g=0,\pi$ are excluded. 

Given those technical tools from Bethe Ansatz and field theory, in this work
we calculate $\rho(n,\t)$ and $\Or(n,\t)$, both for short distances and in the asymptotic limit. We thus obtain the exponents and the amplitudes of the leading uniform and alternating parts and verify them by the Bethe Ansatz results. Interestingly, the leading asymptotics of the alternating part can be directly obtained from those of the uniform part. We furthermore study the limiting values $\t\to0,1-\g/\pi$ in the asymptotic limit, 
where contact is made with $\bra S^z_1S^z_{n}\ket$ and $\langle S^x_1S^x_{n}\ket$.

This article is organized as follows. In the next section, we present the Bethe Ansatz calculation 
of $\rho(n,\t)$ and $\Or(n,\t)$, as well as results from exact diagonalization that we obtained 
additionally. The third part contains the field-theoretical approach. Numerical comparisons 
between the Bethe Ansatz and field-theoretical results are included in an appendix. Calculations not immediately necessary for the understanding of the main text are deferred to further appendices. 
 
\section{Exact short distance string correlation functions }
The Hamiltonian~\refeq{XXZ_Hamiltonian} has been solved exactly by the Bethe Ansatz 
\cite{Bethe31,KorepinBook,TakahashiBook}. In fact the eigenfunctions can be constructed in a form of 
superposition of plane waves, which are called the Bethe Ansatz wave functions. The corresponding eigenenergies 
are obtained by solving the Bethe Ansatz equations 
\be
\te^{\rmi k_j N} = (-1)^{M-1} \prod_{l \ne j} \ds{\frac{\te^{{\rm i} (k_j+k_l)}+1-2 \D e^{\rmi k_j}}
{\te^{\rmi (k_j+k_l)}+1-2 \D \te^{\rmi k_l}}},
\ \ \ \ \ \ \ \ \ \ \ \ \ \ \ \ (j=1,...,M), 
\label{Bethe_Equations} 
\ee 
where ${M}$ is the number of the down spins. With a solution of the Bethe Ansatz equations \refeq{Bethe_Equations}, 
the corresponding eigenenergy is expressed as 
\be
E=\frac{J N  \D}{4} + J \sum_{j=1}^{M} \(\cos k_j - \D \). 
\ee
Especially the ground state is given by a solution in the sector ${M=N/2}$. In the critical 
region ${-1 < \D=\cos\g < 1}$, its value per site in the thermodynamic limit ${N \to \infty}$ 
becomes
\be
e_0 = \frac{J \D}{4} - 
\frac{J \sin \g^2}{4} \int_{-\infty}^{\infty} 
\frac{{\rm d} x }{(\cosh \g x -\cos \g) \cosh \pi x/2} 
\label{Ground_State_Energy_XXZ}
\ee
Enormous works have been contributed to evaluate the physical quantities of the model based on the Bethe 
Ansatz equations~\refeq{Bethe_Equations} \cite{TakahashiBook}. They, however, are usually limited to 
the bulk quantities. Especially, the exact calculation of correlation functions still is a difficult problem. Only for ${\D=0}$, where the system reduces to a lattice free-fermion model 
after a Jordan-Wigner transformation, arbitrary correlation functions can be calculated by means 
of Wick's theorem \cite{Lieb61, McCoy68}. Especially, the two-point spin-spin correlation function is simply 
given by ${\bra S_j^z S_{j+k}^z \ket = -\( 1-(-1)^k \)/(2 \pi^2 k^2)}$.

There have been many attempts to evaluate the correlation functions for general ${\D}$. However, 
explicit exact evaluations of the correlation functions have become attainable only recently. For 
example, the following exact values for the spin-spin correlation functions ${\bra S_j^z S_{j+k}^z \ket}$, were obtained 
up to ${k=7}$ for ${\D=1}$ \cite{Sato05} and up to ${k=8}$  for ${\D=1/2}$ \cite{Kitanine05a}:
\begin{itemize}
\item ${\D=1}$
\be
\zcor{j}{j+1}
&=& \frac{1}{12} - \frac{1}{3}\ln 2
  = -0.147715726853315 \cdots , 
\nonumber \\ 
\zcor{j}{j+2}
&=& \frac{1}{12} - \frac{4}{3} \ln 2 +\frac{3}{4} \zeta(3) 
  =  0.060679769956435 \cdots , 
\nonumber  \\
\zcor{j}{j+3} 
&=& \frac{1}{12} - 3 \ln 2 + \frac{37}{6} \zeta(3) - \frac{14}{3} \ln 2 
\cdot \zeta(3) 
- \frac{3}{2} \zeta(3)^2  \nonumber \\
&& - \frac{125}{24} \zeta(5) + \frac{25}{3} 
\ln 2 \cdot \zeta(5)  = -0.0502486272572352\cdots , \nonumber \\ 
\zcor{j}{j+4} 
&=& \frac{1}{12} - \frac{16}{3}
\ln 2  + \frac{145}{6} \zeta(3) - 54 \ln 2 \cdot \zeta(3) 
- \frac{293}{4} \zeta(3)^2 \nonumber \\
&& - \frac{875}{12} \zeta(5)   + \frac{1450}{3} \ln 2 \cdot \zeta(5)  
- \frac{275}{16} \zeta(3) \cdot \zeta(5) - \frac{1875}{16} \zeta(5)^2 
\nonumber \\
&&+ \frac{3185}{64} \zeta(7)  
- \frac{1715}{4} \ln 2 \cdot \zeta(7) + \frac{6615}{32}  \zeta(3) \cdot \zeta(7)  
\nonumber \\
&=& 0.0346527769827281 \cdots, 
\nonumber \\
\zcor{j}{j+5} &=&-0.0308903666476093\cdots, \nonumber \\
\zcor{j}{j+6} &=& 0.0244467383279589\cdots, \nonumber \\    
\zcor{j}{j+7} &=&-0.0224982227633722\cdots. 
\label{Spin_Correlation_D1}
\ee
\end{itemize}
\begin{itemize}
\item ${\D=1/2}$
\be
\zcor{j}{j+1}&=&-\frac{1}{8} = -0.125,\nonumber \\
\zcor{j}{j+2}&=& \frac{7}{256}=0.02734375, \nonumber \\
\zcor{j}{j+3}&=&-\frac{401}{16384}=0.02447509765625, \nonumber \\
\zcor{j}{j+4}&=&\frac{184453}{16777216} 
=0.0109942555427551\cdots \nonumber \\
\zcor{j}{j+5}&
=&\frac{95214949}{8589934592}
=-0.0110844789305701\cdots, \nonumber \\
\zcor{j}{j+6}&=&\frac{1758750082939}{281474976710656} 
=  0.0062483354772489\cdots, \nonumber \\
\zcor{j}{j+7}&=&-\frac{30283610739677093}{4611686018427387904} 
=-0.0065667113109326\cdots, \nonumber \\
\zcor{j}{j+8}&=&\frac{5020218849740515343761}{1208925819614629174706176} 
=  0.0041526277032786\cdots. 
\label{Spin_Correlation_D05}
\ee
\end{itemize} 
Here ${\zeta(2k+1)}$ is the Riemann's zeta function at odd arguments. Note that the nearest-neighbour correlation function ${\zcor{j}{j+1}}$ can be derived immediately from the ground state energy~\refeq{Ground_State_Energy_XXZ}. So these values have been known long before. We also remark ${\zcor{j}{j+2}}$ for ${\D=1}$ was obtained some decades ago by Takahashi \cite{Takahashi77} by his 
ingenious study of the half-filled Hubbard chain in the strong coupling limit. 
Other results are due to recent developments of the study of the correlation functions.      
Note that even for general ${\D}$, the exact analytic expressions have been obtained up to ${k=3}$ 
\cite{Kato04}. Such progress has enabled comparison with the field-theoretical prediction of the asymptotic behaviour as well as other numerical methods such as numerical diagonalization.

It is interesting to note that the calculation of the spin-spin correlation functions 
Eq.~\refeq{Spin_Correlation_D1} and Eq.~\refeq{Spin_Correlation_D05} rely on the generating function, defined by 
\be
P_n^{\kappa} \equiv
\l\bra
\prod^n_{j=1}
\l\{
\(\frac{1}{2}+S^z_j\)
+\kappa \(\frac{1}{2}-S^z_j\)
\r\}
\r\ket. \label{Spin_Generating_Function} 
\ee
Here ${\kappa}$ is a parameter. Once the generating function Eq.~\refeq{Spin_Generating_Function} is calculated, 
the two-point spin-spin correlation function can be obtained by the formula
\be
\zcor{1}{n}= \frac{1}{2} \frac{\partial^2}{\partial \k^2} 
\left\{ P_n^{\kappa}-2 P_{n-1}^{\kappa}+ P_{n-2}^{\k} \right\} \Big|_{\k=1} 
- \frac{1}{4}. \label{Generarting_Spin_Correlation}
\ee
The generating function (\ref{Spin_Generating_Function}) together with its relation to the 
two-point spin-spin correlation function (\ref{Generarting_Spin_Correlation}) was introduced by Izergin and Korepin \cite{Izergin84, Izergin85}(see also the book \cite{KorepinBook}). Subsequently it was utilized to discuss a certain long-distance asymptotic behaviour \cite{Essler95,Essler96} as well as  to obtain several different forms of multiple integral formulas \cite{Kitanine02,Kitanine05}. However, it was only quite recently the generating function ${P_n^{\kappa}}$ was {\it explicitly} calculated for ${\Delta \ne 0}$, namely, up to ${n=8}$ for ${\D=1}$ \cite{Sato05}  and up to ${n=9}$ for ${\D=1/2}$ \cite{Kitanine05a}.

\begin{figure}[htbp]
  \begin{center}
\leavevmode
  \subfigure[\large ${n=}$ even]{\includegraphics[width=6.5cm]{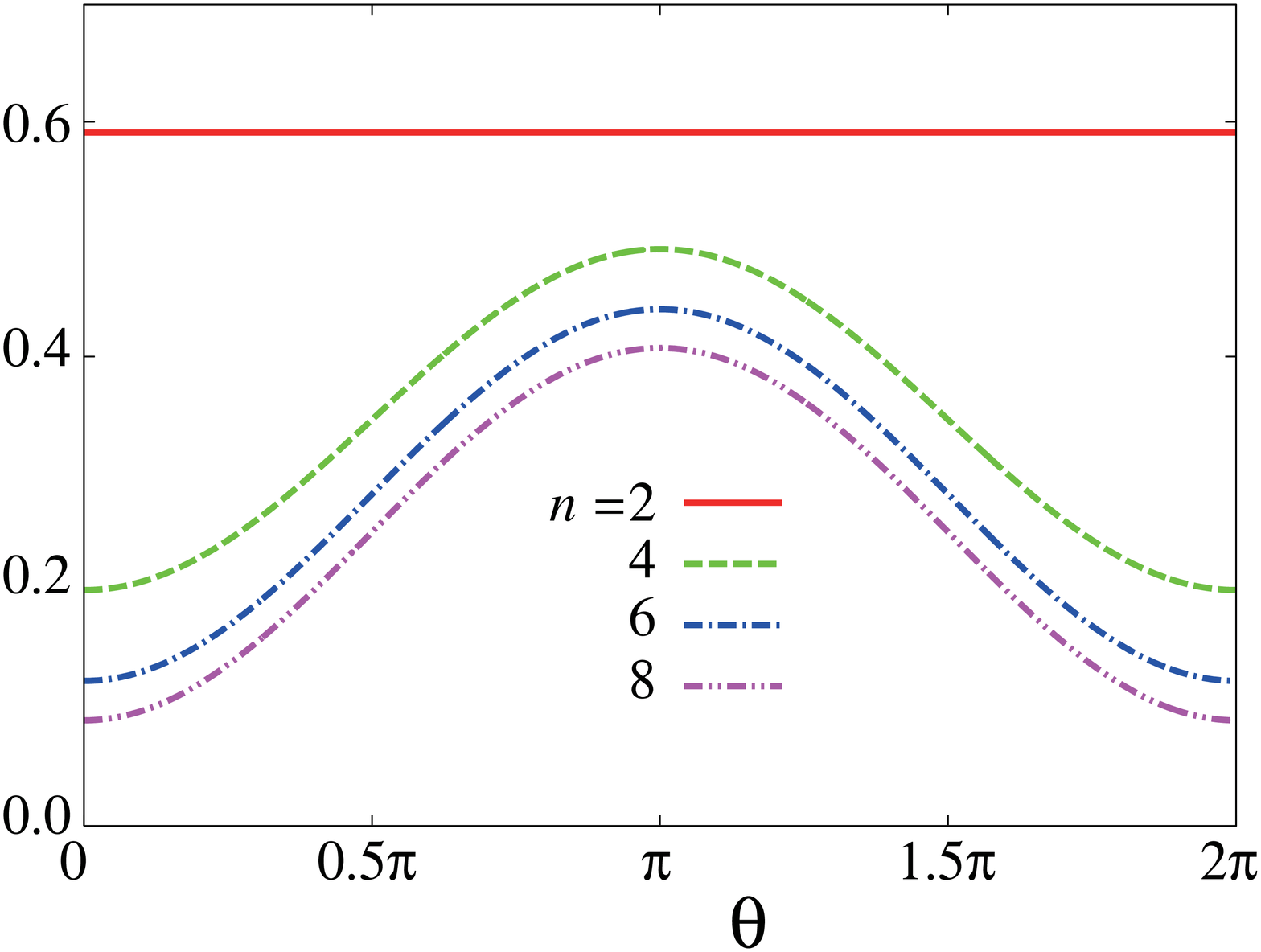}}
  \subfigure[\large ${n=}$ odd]{\includegraphics[width=6.5cm]{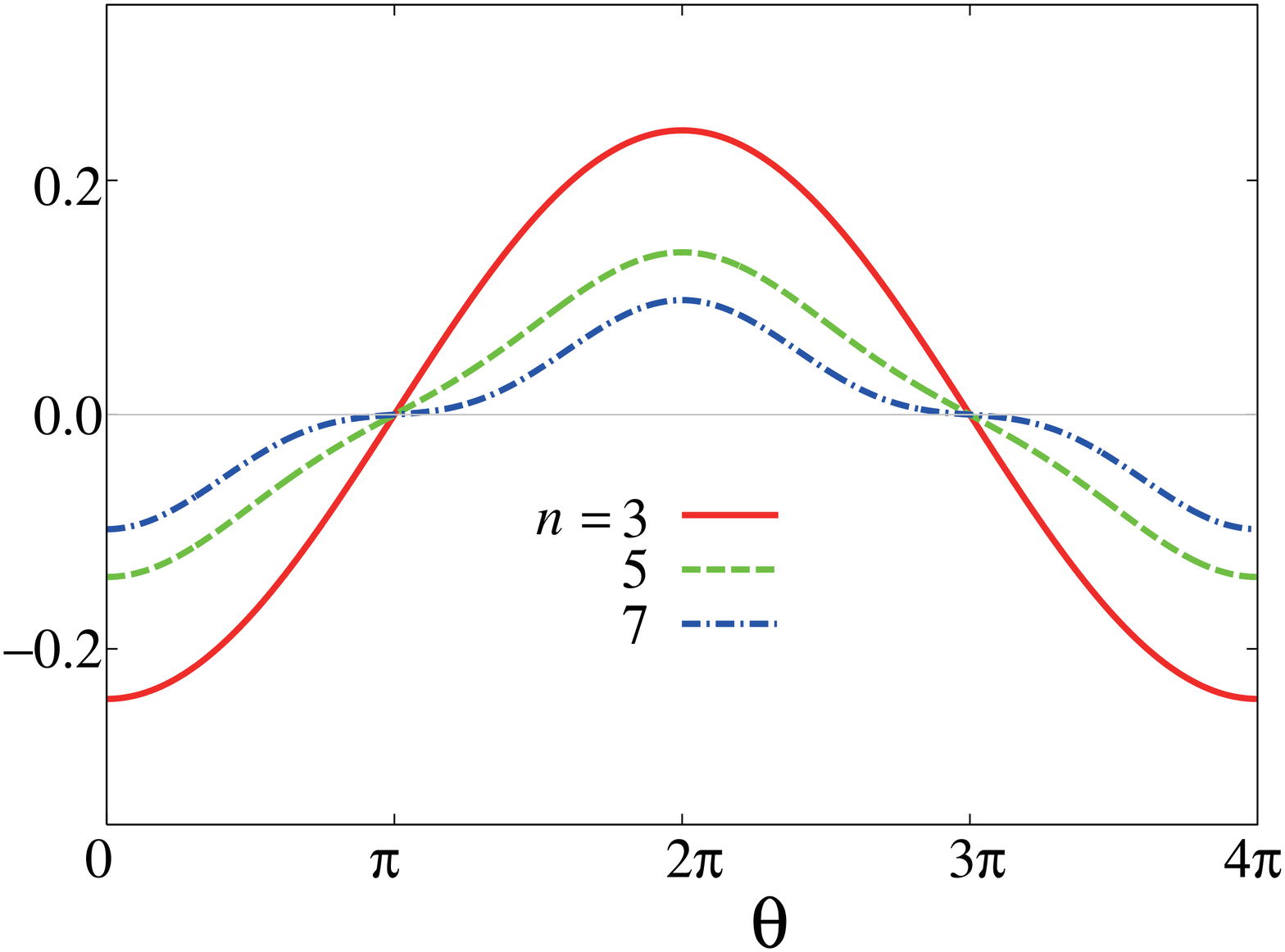}}
    \caption{\large ${\Or(n,\t)}$ for ${\D=1}$}
    \label{figD10}
  \end{center}
\end{figure}

\begin{figure}[htbp]
  \begin{center}
  \subfigure[\large ${n=}$ even]{\includegraphics[width=6.5cm]{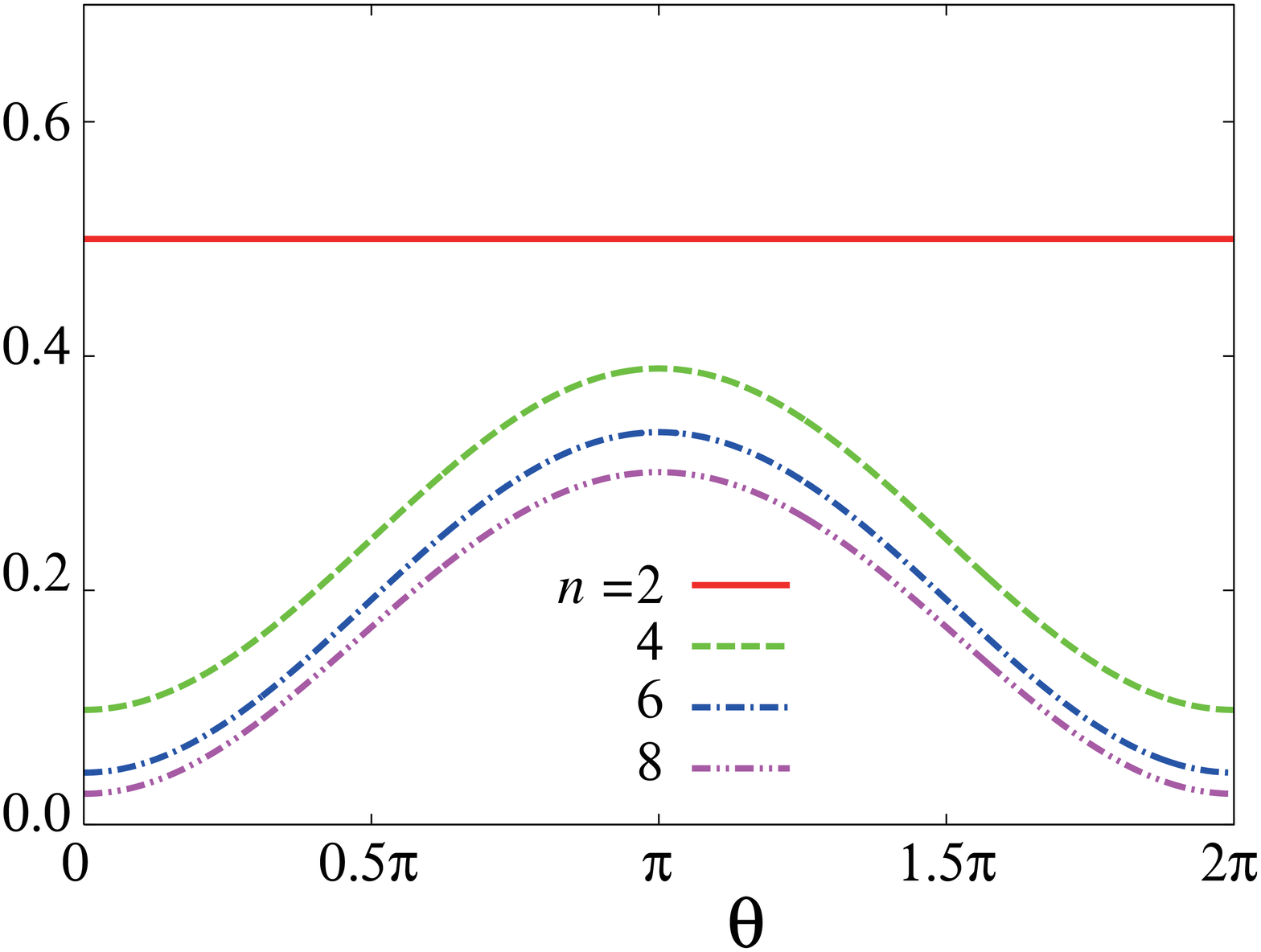}}
  \subfigure[\large ${n=}$ odd]{\includegraphics[width=6.5cm]{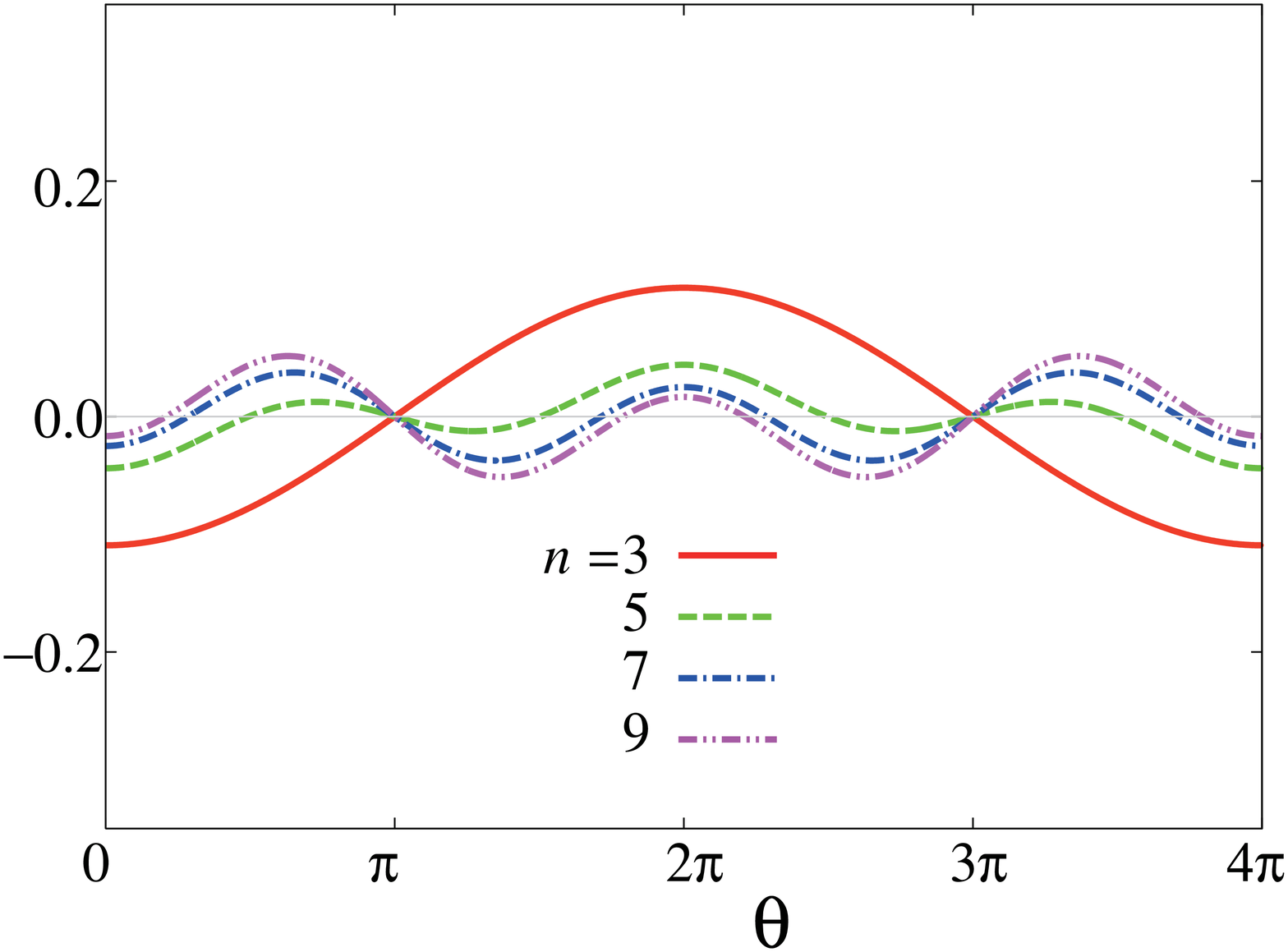}}
    \caption{\large ${\Or(n,\t)}$ for ${\D=1/2}$}
    \label{figD05}
  \end{center}
\end{figure}

Now one will readily find ${P_n^{\kappa}}$, Eq.~\refeq{Spin_Generating_Function} and the string correlation function ${\rho(n,\t)}$, Eq.~\refeq{rho} are connected as
\be
\rho(n,\t) = \k^{-\frac{n}{2}} P_n^{\k} \Big|_{\k=\te^{-\rmi \t}}. 
\ee
Then we can calculate some exact values of ${\rho(n,\t)}$ for ${\D=1}$ and ${\D=1/2}$. 
Moreover, since the generalized string correlation function ${\Or(n,\t)}$
\refeq{ordef} is related to ${\rho(n,\t)}$ as 
\be
\Or(n,\t)&=&\frac{1}{\sin^2\frac\t2}\l[\rho(n,\t) -2\l(\cos\frac\t2\r)
\rho(n-1,\t) + \l(\cos^2\frac\t2\r)\rho(n-2,\t)\r]\label{orfromrho},
\ee
we can also evaluate the generalized correlation functions for ${\D=1}$ and ${\D=1/2}$ 
(cf. Appendix A).  They are plotted in Fig. \ref{figD10} and  Fig. \ref{figD05}. 
From the figures one observes: 
\begin{itemize}
\item For even ${n}$ (${\ge4}$), ${\Or(n,\t)}$ is always positive with a period ${2\pi}$. 
It has a single maximum at ${\t=\pi}$ and a minimum at ${\t=0}$. Recall that ${\Or(n,\pi)=(2 \rmi)^n 
\bra \prod_{k=1}^n S^z_k \ket}$ and ${\Or(n,0)= -4 \bra S^z_1 S^z_n \ket}$.
\item For odd ${n}$, ${\Or(n,\t)}$ has a rather complicated structure with a period ${4\pi}$. 
In this case, ${\Or(n,\pi)}$ and ${\Or(n,3\pi)}$ are  always zero as they should be.  
\end{itemize}
We give some exact values of ${\Or(n,\pi)=(2 \rmi)^n \bra \prod_{k=1}^n S^z_k \ket}$ for ${\D=1}$ 
and ${\D=1/2}$ in the following: 
\begin{itemize} 
\item ${\D=1}$  
\be
\Or(2,\pi) &= & -\frac{1}{3} +\frac{4}{3} \ln 2 = 0.5908629074132604\cdots \nn \\
\Or(4,\pi) &= & \frac{1}{5} - \frac{16}{3} \ln 2 + \frac{232}{15} \zeta(3) - \frac{32}{3} \ln 2 \cdot \zeta(3) 
- \frac{21}{5} \zeta(3)^2 \nn \\ && - \frac{95}{6}\zeta(5) + \frac{70}{3} \ln 2  \cdot \zeta(5) = 0.4914453923615522\cdots \nn \\
\Or(6,\pi) &= & 0.4403016697026268\cdots \nn \\
\Or(8,\pi) &= & 0.4072424147596208\cdots 
\ee
\item ${\D=1/2}$
\be
\Or(2,\pi) &= & \frac{1}{2} \nn \\
\Or(4,\pi) &= & \frac{1595}{4096} = 0.389404296875 \nn \\
\Or(6,\pi) &= & \frac{719423395}{2147483648} = 0.3350076242350041\cdots \nn \\
\Or(8,\pi) &= & \frac{346891287109196331}{1152921504606846976}
           =  0.3008802296800668\cdots
\ee
\end{itemize}
One observes that ${\Or(n,\pi)}$ for ${n}$=even decays very slowly as ${n}$ increases. Namely as mentioned 
in the introduction, for ${\D=1}$, the asymptotic decay ${\Or(n,\pi) \sim n^{-1/4}}$ was given by Hida \cite{Hida92} 
and more generally
\be 
\Or(n,\theta) \sim n^{-\frac{\theta^2}{4 \pi^2}} \label{Lou_Or_Asym}
\ee
by Lou \cite{Lou03}. In the next section, we shall both generalize this asymptotic formula to the more general ${-1< \Delta<1}$ 
case and determine the amplitude by making use of field theory. Furthermore since the formula Eq.~\refeq{Lou_Or_Asym} does not explain the difference of 
the periodicity with respect to the parity of ${n}$, we shall consider some subleading terms more carefully. We remark 
that ${\rho(n,\theta)}$, Eq.~\refeq{rho}, shares periodicity properties analogous to $\mathcal{O}(n,\t)$. 
In fact it is easy to see that ${\rho(n,\t)}$ is expanded as 
\be
\rho(n,\t) = \sum_{j=1}^{n} P_{n,j} \cos\[ \( \frac{n}{2} -j \) \t \], \label{Rho_Expansion}
\ee
where the coefficients ${P_{n,j}}$ are the summation of the diagonal density matrix elements in the sector with ${j}$ 
down spins. Note that ${P_{n,j}=P_{n,n-j}}$. From  Eq.\refeq{Rho_Expansion}, one can immediately find 
\be
\rho(n, \t+2\pi) = (-1)^n \rho(n,\t) \label{Periodicity}
\ee
\noindent
${\bullet}$ ${\D=0}$ case :

Let us comment on the string correlation functions for ${\D=0}$. In this case, a simple 
determinant formula for the string correlation function ${\rho(n,\theta)}$ exists (c.f. Ref.\cite{Colomo93})
\be
\rho(n,\t) = \det\[ \(\cos \frac{\theta}{2} \)\delta_{j,k} +  \(\rmi \sin \frac{\theta}{2} \right) M_{j,k} \]_{j,k=1}^{n},
 \ \ \ \ M_{j,k} =
\begin{cases} 0 & j-k:{\rm even} \\ 
           \dfrac{2}{\pi} \dfrac{(-1)^{\frac{j-k+1}{2}}}{j-k} & j-k : {\rm odd}
\end{cases} \nonumber \\
\label{RhoD0}
\ee
Using this formula one can evaluate the exact numerical values up to the order of ${n \simeq 10000}$ easily, for example, by {\it Mathematica} on a standard PC. We give the exact values in Table \ref{exactDelta0} up to ${n=1000}$.

\begin{table}[h]
\begin{center}
\begin{small}
\begin{tabular}
{@{\hspace{\tabcolsep}\extracolsep{\fill}}ccccccccc} 
\hline 
$n$ & 5 & 10 & 20 & 50 & 100 & 200 & 500 & 1000 \\ 
\hline
$\rho(n,\pi/4)$ 
& 0.884857 & 0.866761 & 0.848076 & 0.824090 & 0.806421 
& 0.789137 & 0.766860 & 0.750428  \\
$\rho(n,\pi/2)$ 
& 0.605461 & 0.564975 & 0.516684 & 0.459997 & 0.421580 
& 0.386481 & 0.344597 & 0.315979 \\
$\rho(n,3 \pi/4)$ 
& 0.289075 & 0.291125 & 0.234367 & 0.177503 & 0.144555 
& 0.118072 & 0.0906455 & 0.0743400  \\
\hline
\end{tabular}
\end{small}
\end{center}
\caption{Exact values of ${\rho(n,\theta)}$ for ${\D=0}$}
\label{exactDelta0}
\end{table}
This determinant is also expressed as a Toeplitz determinant
\be
\rho(n,\t) &=& \te^{\rmi n \t/2} \det\l[ \dfrac{1}{2 \pi} \int_{0}^{2 \pi} {\rm e}^{{\rm i} (j-k) q} \sigma(q) {\rm d} q  \right]_{j,k=1}^{n}, \ \sigma(q) =
\begin{cases} 
{\rm e}^{-{\rm i} \theta} & : 0 < q < \dfrac{\pi}{2} \nonumber \\ 
               1 &    :  \ \dfrac{\pi}{2} < q < \dfrac{3 \pi}{2} \nonumber \\
               {\rm e}^{-{\rm i} \theta} & : \dfrac{3\pi}{2} < q < 2\pi  \\ 
\end{cases} 
\\
\ee
There are some mathematical results known on the asymptotic behaviours of Toeplitz determinants as ${n \to \infty}$. Assume ${\theta \ne 0, 2 \pi}$, then  ``the generating function" ${\sigma(q)}$ of the Toeplitz determinant has jump singularities at ${q=\pi/2}$ and ${q= 3 \pi/2}$. In such a case, we can invoke the (generalized) Fisher-Hartwig conjecture \cite{Fisher68,Basor91}, which brings about an asymptotic formula for ${0< \theta <2 \pi}$ as
\be
\rho(n, \theta)  &\simeq &  \rho_{\rm Asym}^{(0)}(n,\theta) + (-1)^n \rho_{\rm Asym}^{(1)}(n,\theta) , \nn \\
\rho_{\rm Asym}^{(k)}(n,\theta) & =& n^{-2 (- \frac{\theta}{2 \pi}+k)^2} 4^{- (- \frac{\theta}{2 \pi}+k)^2} 
\left[G(1+\frac{\theta}{2 \pi}-k)
G(1-\frac{\theta}{2 \pi}+k)\right]^2. \label{AsymRhoDelta0}
\ee
Here ${G(z)}$ is the Barnes $G$-function defined by 
\be
G(z+1)=(2 \pi)^{\frac{1}{2}z} 
\exp \left(-\frac{1}{2}z - \frac{1}{2}(\gamma+1) z^2\right)
\prod_{k=1}^{\infty} \left\{ \left(1+\frac{z}{k} \right)^k 
\exp\left(-z+\frac{z^2}{2 k} \right) \right\} \nonumber  
\ee
where ${\gamma=0.5772156649\cdots}$ is the Euler-Mascheroni constant. 
 Each term ${\rho_{\rm Asym}^{(k)}(n,\theta)}$ decays algebraically with the exponent ${-2\left(- \frac{\theta}{2 \pi} +k \right)^2}$. Then the dominant term is ${\rho_{\rm Asym}^{(0)}(n,\theta)}$ for ${ 0 <\theta< \pi}$, and  ${(-1)^n \rho_{\rm Asym}^{(1)}(n,\theta)}$ for ${ \pi<\theta< 2 \pi}$. We refer the reader also to Refs.~\cite{Jin04,Franchini05} for more information about the (generalized) Fisher-Hartwig conjectures. 
\begin{table}[h]
\begin{center}
\begin{small}
\begin{tabular}
{@{\hspace{\tabcolsep}\extracolsep{\fill}}ccccccccc} 
\hline
 $n$  & 5 & 10 & 20 & 50 & 100 & 200 & 500  & 1000 \\ 
\hline
$\rho_{\rm Asym}(n,\pi/4)$ 
& 0.884970 & 0.866783 & 0.848081 & 0.824091 & 0.806421 
& 0.789137 & 0.766860 & 0.750428 \\
$\rho_{\rm Asym}(n,\pi/2)$ 
& 0.605720 & 0.565076 & 0.516705 & 0.460000 & 0.421581 
& 0.386481 & 0.344597 & 0.315979     \\
$\rho_{\rm Asym}(n,3 \pi/4)$ 
& 0.289328 & 0.291316 & 0.234403 & 0.177507 & 0.144555 
& 0.118072 & 0.0906455 & 0.0743400  \\
\hline
\end{tabular}
\end{small}
\end{center}
\caption{Numerical values obtained from the asymptotic formula ${\rho_{\rm Asym}(n,\theta)}$ 
for ${\D=0}$}
\label{AsymDelta0}
\end{table}

Numerical values calculated from Eq.~\refeq{AsymRhoDelta0} are listed in Table \ref{AsymDelta0}. 
Good agreement is found with the data in Table \ref{exactDelta0}. In this context, it is remarkable that they coincide within at least three digits even for small sizes as ${n=10}$. 
Finally let us note a further exact result for ${\rho(n,\pi)}$ at ${\D=0}$. Since ${\rho(2m-1,\pi)=0}$, we consider ${\rho(2m,\pi)}$, which is given more explicitly as
\be
\rho(2m,\pi) &=& (-1)^m 2^{2m} \l\bra \prod_{j=1}^{2m} S_j^z \r\ket =  (-1)^m  {\rm det} \left[ M_{j,k}\right]_{j,k=1}^{2 m} \nn \\
             &=&  \left( \frac{2}{\pi} \right)^{2m} 
             \prod_{k=1}^{m} \prod_{j\ne k}^m \left( \frac{j-k}{j-k-1/2} \right)^2 = \prod_{k=1}^{m} \frac{\Gamma^4(k)}{\Gamma^2(k-\frac{1}{2})    
             \Gamma^2(k+\frac{1}{2})} \nn \\
             &=& \exp \left[ -\frac{1}{2} \int_{0}^{\infty} \frac{{\rm d} t}{t} \frac{1 - {\rm e}^{- mt}}{\cosh^2(t/4)} \right] \nn \\
             &=&c_0 m^{-1/2} \exp \left[ -\frac{1}{2} \int_{0}^{\infty} \frac{{\rm d} t}{t}  {\rm e}^{- t}\tanh^2 \left(\frac{t}{4m}\right) \right] \nn \\ 
             &=&c_0 m^{-1/2} \left( 1 -\frac{1}{32} m^{-2} + \frac{17}{2048} m^{-4}- \frac{379}{65536} m^{-6}+ \cdots \right),  \label{rhoff}           
\ee
where 
\be
c_0&=&{\rm exp} \left[ \frac{1}{2} \int_{0}^{\infty} \frac{{\rm d} t}{t} \left( {\rm e}^{-4t} - \frac{1}{\cosh^2 t} \right)\right] = \left[ G\left(\frac{1}{2}\right) G\left(\frac{3}{2}\right) \right]^2.\nn
\ee
Here we have used an integral formula for the logarithm of the Euler gamma function,
\be
\log \Gamma(z)
=\ip \left[ (z-1) - \frac{1-{\rm e}^{(-z+1)t}}{1-{\rm e}^{-t}} \right]\frac{{\rm e}^{-t}}{t} {\rm d}t, \ \ \ \ \ \ (\Re(z)>0).
\ee
Thus we can obtain the asymptotic expansion to an arbitrary order in this case.  Note that the leading term is consistent with the formula Eq.~\refeq{AsymRhoDelta0} with ${\theta=\pi}$. 
\section{Asymptotic behaviour of string correlation functions}
\setcounter{equation}{0}

In this section, we will discuss the asymptotic behaviour of the string correlation functions for the critical region ${-1<\Delta<1}$ (that is $\pi>\g>0$) by use of field theoretical arguments. Thus the aim is to find coefficients $D_j$ and exponents $\nu_j$ such that
\be
\lim_{n\to\infty} \frac{\rho(n,\t)-\sum_{j=1}^{m-1} D_j(\t,\g) n^{-\nu_j(\t,\g)}}{n^{-\nu_{m}(\t,\g)}}=:D_{m}(\t,\g) \;\, \text{(finite)},\;\, m=1,2,\ldots \label{asdef}.
\ee
The exponents are increasing with $j$, i.e. $\nu_j<\nu_{j+1}$. The amplitudes and exponents depend on the parameters $\t,\g$ of the model and of the function $\rho$. Instead of Eq.~\refeq{asdef}, we use the shorthand notation
\be
\rho(n,\t)\sim \sum_j D_j(\t,\g) n^{-\nu_j(\t,\g)}\nn.
\ee
The important point to remember is that the asymptotic expansion is defined in the limit $n\to \infty$.  

We first present the results obtained so far within the field-theoretical framework and give details of the derivation in the following section. 

  \subsection{Results}
  We find the following asymptotic expansion of the string correlation
  function for $0< \t\leq \pi$:
  \be
  \rho(n,\theta)&\equiv& \l\langle \exp\l[\rmi \theta \sum_{k=1}^n
    S_k^z\r]\r\rangle\nn\\
  &\sim& D(\theta,\g)n^{-\nu_1(\theta,\g)}\l(1+\mathcal O\l( n^{-\delta(\g)}\r)\r)\nn\\
  & &+ (-1)^n D(2 \pi-\theta,\g)
  n^{-\nu_1(2\pi-\theta,\g)} \l(1+\mathcal O\l(n^{-\delta(\g)}\r)\r)\label{ras}\\
  & & +\mathcal O\l(n^{-\nu_1(\theta,\g)-2},(-1)^n n^{-\nu_1(2\pi+\theta,\g)},(-1)^n n^{-\nu_1(2\pi+\theta,\g)-2},(-1)^n n^{-\nu_1(2\pi-\theta,\g)-2}\r)\, ,\nn
  \ee
  with
  \be
  \nu_1(\t,\g)&=&\frac{\t^2}{4\pi^2}\,\frac{\pi}{\pi-\g}\nn\\
  \delta(\g)&=&4 \frac{\pi}{\pi-\g} -4\nn\\
  D(\t,\g)&=& \l[\frac{\G\l(\frac{\eta}{2-2\eta}\r)}{2\sqrt{\pi}
      \G\l(\frac{1}{2-2\eta}\r)}\r]^{\t^2/(4\eta\pi^2)}\nn\\
  & &\times \exp\l[-\ip
    \l(\frac{\sinh^2\frac{\theta}{2\pi} t}{\sinh t \,\cosh(1-\eta) t\,\sinh
      \eta t} - \frac{\t^2 \te^{-2 t}}{4\eta \pi^2}\r)\frac{\d t}{t}\r]\label{coeffluk}\\
  &=&\l[\frac{\G\l(\frac{\pi R^2}{1-2\pi R^2}\r)}{2\sqrt{\pi}
      \G\l(\frac{1}{2-4\pi R^2}\r)}\r]^{(\t/(2\pi R))^2/(2\pi)}\nn\\
  & & \times \exp\l[-\ip
    \l(\frac{\sinh^2\frac{\theta}{2\pi} t}{\sinh t \,\cosh(1-2\pi R^2) t\,\sinh
      2\pi R^2 t} - \l(\frac{\t}{2\pi R}\r)^2\frac{ \te^{-2 t}}{2\pi}\r)\frac{\d t}{t}\r] \nn \\ \label{coeffr}
  \ee
  In Eq.~\refeq{coeffluk}, Lukyanov's notation is used with
  $\eta=\frac{\pi-\g}{\pi}$, whereas in Eq.~\refeq{coeffr}, the anisotropy
  is written in terms of the compactification radius $R$ with $2\pi
  R^2=\eta$. 
  
  Since $\rho(n,\t)=\rho(n,-\t)$, the result \refeq{ras} is readily extended
  to the domain $-\pi\leq \t < 0$. Thus $\rho(n,\t)$ is known in the
  fundamental domain $-\pi\leq \t\leq \pi$ (note ${\rho(n,\t=0)=1}$, trivially). 
  The periodicity Eq.~\refeq{Periodicity} then yields $\rho$ for all values of $\t$. 
  
  We note the following limiting values of the coefficient $D(\t,\g)$: 
  \begin{itemize}
  \item $D(\t=2\pi \eta,\g)=2 (1-\eta)^2 A$, where $A$ is the
    coefficient of the leading term in an asymptotic expansion of 
    the uniform part of $\langle
    \sigma^x_{1} \sigma^x_n\rangle$, namely: $\langle
    \sigma^x_{1} \sigma^x_n\rangle_u\sim \frac{A}{n^\eta}$,
    \cite{Lukyanov03}. Then, as ${\nu_1(\t=2\pi\eta,\g)=\eta}$, we have the
    asymptotic equality  (note that $1-\eta=\g/\pi$) for $\pi/2\leq \g <\pi$
    \be
    \rho(n,2\pi \eta)\sim 2\l(\frac{\g}{\pi}\r)^2\langle
    \sigma^x_{1} \sigma^x_n\rangle_u,\ \ \ \ \pi/2\leq \g <\pi\label{xx}
    \ee
    for the leading order of the uniform part (in order to facilitate
    comparison with Lukyanov's results, we use the Pauli-matrices
    $\sigma^\nu=2 S^\nu$). This is checked immediately at the free fermion point $\D=0$ from Eq.~\refeq{rhoff}.
  \item $D(0,\g)=1$, whereas
    \be
    \lim_{\t\to0}D(2\pi-\t,\g) \frac{16}{\t^2} &=&\frac{A_z}{2}\nn\\
    &\equiv& \frac{4}{\pi^2}
    \l[\frac{\G\l(\frac{\eta}{2-2\eta}\r)}{2\sqrt{\pi}
	\G\l(\frac{1}{2-2\eta}\r)}\r]^{1/\eta} \nn\\
    & & \times\exp\l[ \ip
      \l(\frac{\sinh((2\eta-1) t)}{\sinh(\eta t) \cosh((1-\eta) t)}-\frac{2
	\eta-1}{\eta} \te^{-2 t}\r)\frac{\d t}{t}\r] \nn \\ \label{limit}.
    \ee
    The last equation is proved in Appendix B. Following Lukyanov's notation \cite{Lukyanov03}, $A_z$ denotes the
    coefficient of the leading contribution in the alternating part $\langle \sigma_1^z\sigma_n^z\rangle_a$ of the
    $\sigma^z$-$\sigma^z$-correlation function, namely
    \be
    \langle \sigma^z_{1} \sigma_n^z\rangle_a 
    \sim\frac{(-1)^{n-1} A_z}{n^{1/\eta}}
    \label{szsza}.
    \ee
  \end{itemize}
  In order to obtain the asymptotics of the generalized string correlation
  function, we first express it in terms of $\rho(n,\t)$ according to Eq.~\refeq{orfromrho}. 
  Then, using the above results, the asymptotic behavior of $\Or(n,\t)$ is obtained
  for ${0<\theta \le\pi}$:
  \be
  &&\Or(n,\theta) \equiv -4\l \langle S_1^z \exp\l[\rmi
    \theta \sum_{k=2}^{n-1} S_k^z\r] S_n^z\r\rangle\nn\\
  &\sim&
  D(\t,\g) n^{-\nu_1(\t,\g)} \l[\tan^2 \frac\t4 -\frac{\cos\frac\t2}{\cos^2\frac\t4}\,\frac{\nu_1(\t,\g)}{n}+
  \frac{\cos\frac\t2(2\cos\frac\t2-1)}{\sin^2\frac\t2}\frac{\nu_1(\t,\g)(\nu_1(\t,\g)+1)}{n^2} +\cdots \r] \nn\\
  && + (-1)^n D(2\pi-\t,\g) n^{-\nu_1(2\pi-\t,\g)} \nn \\ 
  && \times \Bigg[\cot^2 \frac\t4 + \frac{\cos\frac\t2}{\sin^2\frac\t4}\,
  \frac{\nu_1(2\pi-\t,\g)}{n} + \frac{\cos\frac\t2(2\cos\frac\t2+1)}{\sin^2\frac\t2}\frac{\nu_1(2\pi-\t,\g)(\nu_1(2\pi-\t,\g)+1)}{n^2} +\cdots \Bigg]  \nn\\
  \label{oasym}
  \ee
Let us consider the limit $\t\to 0$ of the asymptotic formula Eq.\eqref{oasym}. 
The first two terms of the uniform part in \eqref{oasym} vanish in this limit and the third term gives 
\be
\lim_{\t\to0} D(\t,\g) n^{-\nu_1(\t,\g)} \l[  \frac{\cos\frac\t2(2\cos\frac\t2-1)}{\sin^2\frac\t2}\frac{\nu_1(\t,\g)(\nu_1(\t,\g)+1)}{n^2} \r] =\frac{1}{\pi(\pi-\g)n^2}=\frac{1}{\pi^2 \eta n^2}
\label{limit1}
\ee
Because of the relation Eq.~\eqref{limit}, the leading alternating part yields 
\be
\lim_{\t\to0}  (-1)^n D(2\pi-\t,\g) n^{-\nu_1(2\pi-\t,\g)} \cot^2 \frac\t4 =  \frac{(-1)^n A_z}{2 n^{1/\eta}}
\label{limit2}
\ee
In order to get the correct leading alternating term in the limit $\t\to 0$, we should also consider the leading alternating part for ${-\pi\le\t<0}$, which reads
\be
(-1)^n D(2\pi+\t,\g) n^{-\nu_1(2\pi+\t,\g)} \l[\cot^2 \frac\t4 + \frac{\cos\frac\t2}{\sin^2\frac\t4}\,
  \frac{\nu_1(2\pi+\t,\g)}{n} +\cdots\r] 
\ee
in addition to \eqref{oasym}. Then we have similarly to Eq.~\eqref{limit2}
\be
\lim_{\t\to0}  (-1)^n D(2\pi+\t,\g) n^{-\nu_1(2\pi+\t,\g)} \cot^2 \frac\t4 =  \frac{(-1)^n A_z}{2 n^{1/\eta}}
\label{limit3}
\ee
in the limit ${\t\to0}$. Collecting the terms \eqref{limit1}, \eqref{limit2} and \eqref{limit3} yields
\be
\lim_{\t\to0} \Or(n,\t) \sim \frac{(-1)^n A_z}{ n^{1/\eta}} + \frac{1}{\pi^2 \eta n^2} .
\label{limitOr}
\ee
Eq.~\refeq{limitOr} coincides with the leading asymptotic behaviour of two-point correlation function 
${-4\l\langle S_1^z S_n^z\r\rangle=-\l\langle \sigma_1^z \sigma_n^z\r\rangle}$.


  \subsection{Derivation}
  \subsubsection{The string correlation function $\rho(n,\t)$}
  An effective field theory describing the low-energy-excitations of the
  XXZ-chain in the critical regime $0<\g<\pi$ has been derived by Lukyanov
  \cite{Lukyanov98}. At zero magnetic field, the corresponding
  Hamiltonian density $\mathcal H$ reads
  \be
  \mathcal H&=& \frac{v}{2} \l(\Pi(x)^2+(\partial_x\varphi(x))^2\r) +\ve^{1/(\pi R^2)-2}\la \cos
  \frac{2\vp(x)}{R} \nn\\
  & &+ \ve^{2}\l[\la_+ J_L^2(x) J_R^2(x) +
  \la_-\l[J_L^4(x)+J_R^4(x)\r]\r] +\ldots\label{ham} ,
  \ee
  where the dots denote terms with scaling dimensions higher than those given
  explicitly. The dimensionless constants $v,\la,\la_\pm$ are known exactly from Bethe Ansatz 
  \cite{Lukyanov98}. The lattice constant $\ve$ has the dimension of a length, whereas the dimension
  of $\mathcal H$ is 1/length$^2$. Above, $\mathcal H$ is the sum of a
  Gaussian model and irrelevant operators, the latter with scaling dimensions
  $1/(\pi R^2)$ and $4$. The Hamiltonian is written in terms of the bosonic
  field $\varphi$ and its momentum $\Pi$, where $\l[\varphi(x),\Pi(y)\r]=\rmi
  \delta(x-y)$. The left- and right current operators are then defined as
  $J_{L,R}(x)=\frac{\mp 1}{\sqrt{4\pi}}\l(\Pi(x)\pm \vp(x)\r)$. 

  Within the same approach, the effective $S^z$-operator reads
  \be
  S^z_j\equiv S^z(x)& \sim& \frac{\ve}{2\pi R} \partial_x\vp(x) + \sum_{m=0}^\infty (-1)^m
  \ve^{(2m+1)^2/(4\pi R^2)}C_m \cos
  \l(\frac{2 m+1}{R} \vp(x)\r)\nn\\
& &+ \mbox{descendants}\,,\label{szas}
  \ee
  where $x=\ve j$. The constants $C_m$ have been determined in
  \cite{Lukyanov03}. We do not write down the descendant fields here explicitly, but only 
  note that if a primary field has scaling dimension $\Delta$, then the descendant 
  fields have scaling dimension $\Delta+\ell$, where $\ell$ is a certain positive integer.

  To arrive at an asymptotic expression for $\rho(n,\t)$, we first apply the
  Euler-MacLaurin formula to the sum in the exponent:
  \be
  \sum_{k=1}^n S_k^z &=&\ve^{-1}\int_0^x S^z(x') \d x'-\frac12
  \l[S^z(0)-S^z(x)\r]+ \Or(\partial^\mu_x S^z)\nn\\
  &=&\frac{1}{2\pi R} \l(\vp(x)-\vp(0)\r) + \mathcal O(\ve^{(2m+1)^2/(4\pi R^2)+\mu},\ve^\mu)\label{sumsz},
  \ee
  ($\mu\geq 1$ integer) from which one concludes that the only cutoff-independent contribution in
  the integral stems from the first term in the last equation. We expand the
  exponent with respect to $\ve$ and arrive at
  \be
  \rho(n,\t)\sim \l\langle \exp\l[\rmi \frac{\t}{2\pi
  R}(\vp(x)-\vp(0))\r]\l(1+\mathcal O(\ve^{2k\l[(2m+1)^2/(4\pi
  R^2)+\mu\r]},\ve^{2k\mu})\r)\r\rangle
  \ee
  where the positive integer $k$ originates in the expansion of the exponential
  function. From this we conclude that the leading exponent of the uniform
  part is $\nu_1(\t,\g)$,
  and the leading Euler-MacLaurin corrections to this have exponents $\nu_1(\t,\g)+2,
  \nu_1(\t,\g)+1/(2\pi R^2)$. 

  Thus in order to determine the amplitude of the leading term, we have to calculate\\
  $\l\langle \exp\l[\rmi \frac{\t}{2\pi
  R}(\vp(x)-\vp(0))\r]\r\rangle$. In the field-theoretical setting of massless
  Bose fields considered here, this quantity is defined only up to a
  multiplicative constant $\L$ with dimension 1/length \cite{Lukyanov03}. It has
  become custom to choose it such that (``CFT normalization condition'')
  \be
  \L^{\al^2/(2\pi)}\langle\exp\l[\rmi \al(\vp(x)-\vp(0))\r]\rangle=|x|^{-\al^2/(2\pi)}.
  \ee
  This means that we have to introduce a constant $D(\t,\g)$ as follows:
  \be
  \rho(n,\t)\sim D(\t,\g) \l\langle \exp\l[\rmi \frac{\t}{2\pi
      R}(\vp(x)-\vp(0))\r]\r\rangle=\frac{D(\t,\g)}{n^{(\t/2\pi R)^2/(2\pi)}}\label{lead}
  \ee
  for the leading decay of the uniform part. Because of the symmetry $\rho(n,-\t)=\rho(n,\t)$, 
  this result is valid for
  $-\pi\leq \t\leq \pi$. Let us defer the calculation of
  the coefficient $D$ to the next paragraph and first determine the leading
  exponent of the alternating part. This is obtained directly by exploiting
  the periodicity Eq.~\refeq{Periodicity}. Together with Eq.~\refeq{lead}, it
  implies that
  \be
  \rho(n,\pi\leq \t\leq 3\pi) &\sim& D(\t-2\pi,\g) n^{-\nu_1(\t-2\pi,\g)}\label{nlead1}\\
  \rho(n,-3\pi\leq \t\leq -\pi) &\sim& D(\t+2\pi,\g) n^{-\nu_1(\t+2\pi,\g)}\label{nlead2}.
  \ee
  The exponents are expected to depend continuously on the parameters
  $\t,\g$. Thus for $0\leq \theta\leq \pi$ ($-\pi\leq\t<0$), Eq.~\refeq{nlead1}
  (Eq.~\refeq{nlead2}) yields the leading contribution to the alternating
  part, which is next-leading with respect to the leading decay given in
  Eq.~\refeq{lead}. 

  What are the exponents of the next-leading contributions? In
  Eq.~\refeq{lead} we have tacitly assumed that the expectation value is taken
  with respect to the unperturbed Gaussian part of the Hamiltonian
  \refeq{ham}. However, there are additional contributions in Eq.~\refeq{ham},
  with scaling dimensions $\D=1/(\pi R^2),4$. As argued in \cite{Zamolodchikov91}, they
  lead to exponents $\nu_1(\t,\g)+ k( \D-2)$ in $\l\langle \exp\l[\rmi \frac{\t}{2\pi
      R}(\vp(x)-\vp(0))\r]\r\rangle$, where the integer $k$ denotes the order of the
  perturbational expansion. Since there is no contribution of the
  $\cos$-operator for $k=1$, the
  next-leading exponent stemming from this contribution is
  $\nu_2(\t,\g)=\nu_1(\t,\g)+\delta(\g)$ with $\delta(\g)=4\pi/(\pi-\g)-4$.  On the other hand, the first-order
  contribution of the $\la_\pm$-operators yields an exponent
  $\nu_1(\t,\g)+2$. This latter one is always larger than $\nu_1(\t,\g),
  \nu_1(\t-2\pi,\g)$ (for $0<\t<\pi$) and we discard it here. Thus $\nu_2(\theta,\gamma)$
  yields the next-leading exponent in the uniform part. According to the periodicity argument,
  the next-leading exponent in the alternating part for $0<\t<\pi$ is then $\nu_2(\t-2\pi,\g)$.

  We now focus on the coefficient $D(\t,\g)$. The result given in
  Eq.~\refeq{coeffluk} is a conjecture based on the work \cite{Lukyanov97}. 
The following tests of this conjecture have been performed:
\begin{itemize}
  \item For $\g=\pi/2$,  one can show that $D(\theta,\pi/2)$ reduces to (\ref{AsymRhoDelta0}), namely 

\be
    D(\theta,\pi/2)&=& 4^{-\frac{\t^2}{4\pi^2}}
    \l[G\l(1+\frac{\t}{2\pi}\r)G\l(1-\frac{\t}{2\pi}\r)\r]^2.
    \label{coeffDelta0}
\ee
This equality can be checked by means of an integral representation of the Barnes $G$-function (see Appendix C).
\item Numerical comparisons for $\Delta=1/2$ between the exact data from the
    Bethe Ansatz (for $n=9$) and the asymptotic results \eqref{ras},\eqref{oasym} have been performed for $\t=\pi/4,\pi/2,3\pi/4,\pi$. In all cases, very good agreement is found. Similarly, we compared with the data obtained by the numerical diagonalization up to a system size of ${N=28}$ lattice sites 
    for general ${\Delta}$ (see Appendix A).  
\end{itemize}
  Our conjecture for $D$ is based on arguments similar to the conjecture for the coefficient of the leading 
  decay of $\langle \sigma^x_{1}
  \sigma^x_n\rangle$, cf. \cite{Lukyanov03}, \cite{Lukyanov97}. In \cite{Lukyanov97}, the
  expectation value of $\langle\exp\l[\rmi \alpha \vt\r]\rangle$ in a {\em
    massive} Sine-Gordon model with an operator $\cos(\beta\vt)$ is determined,
  \be
  \langle\exp\l[\rmi \alpha \vt\r]\rangle=(\ve m)^{\alpha^2/(4\pi)} N(\alpha,\beta)\label{expal},
  \ee
  where $m$ is the particle mass associated with the field $\vt$. Since in that problem, $\sigma^x\sim \te^{\rmi 2\pi R \vt}$ with an a priori unknown amplitude, calculating the amplitude of the leading decay of $\langle \sigma^x_{1}
  \sigma^x_n\rangle$  with respect to a Sine-Gordon model with an operator
  $\cos(\beta\vt)$ is very similar to our problem of determining $D$. 

  An explicit value for $N(\al,\beta)$ in Eq.~\refeq{expal} is conjectured and confirmed explicitly
  in certain limiting cases in \cite{Lukyanov97}. The authors then calculate $\langle
  \sigma_{1}^x \sigma_n^x\rangle\sim A(\eta) N(1/\eta,2/R) n^{-\eta}$ by making use of the fact
  that this correlation function is known {\em explicitly} for the massive
  XYZ-model close to the critical XXZ-point, namely  $\langle
  \sigma_{1}^x \sigma_n^x\rangle_m\sim A_m  (\ve m)^{-\eta}$ with
  a known coefficient $A_m$. This allows for the deduction of $A(\eta)$. 

  In our case, the field $\vp$ is related to $\vt$ by
  $\partial_t\vp=\partial_x\vt$. However, the problem of calculating $D$ is
  completely analogous to the calculation of $A(\eta)$ sketched above, with a
  Sine-Gordon-term $\cos(2\vp/R)$ in the Hamiltonian. The
  only unknown is the string function $\rho_m(\t)$ in the massive
  XYZ-regime. We know that 
  \be
  \rho_m(\t)= C_m(R) (\ve m)^{(\t/(2\pi
  R))^2/(2\pi)}\label{massrho}
  \ee
  with an unknown constant $C_m$ depending on $R$. Note that
  in the massive regime, we cannot relate $2/R$ to $\g$, but rather take it as
  the constant in the Sine-Gordon-term $\cos(2\vp/R)$. On the other hand, the
  results in \cite{Lukyanov97} tell us that
  \be
  \rho_m(\t)=D\, N(\theta/(2\pi R), 2/R) (\ve m)^{(\t/(2\pi R))^2/(2\pi)}\label{massrho2}
  \ee
  with a known coefficient $N(\theta/(2\pi R),2/R)$. By comparing
  Eq.~\refeq{massrho2} with Eq.~\refeq{massrho}, one obtains $D$ in terms of
  $\theta$, $R$ and the unknown $C_m(R)$. We find that
  $C_m(R)=2(1-\eta)^2=2(1-2\pi R^2)^2$ yields excellent agreement with the
  numerical data as described above. This results in the coefficient
  $D(\t,\g)$ as given in Eqs.~\refeq{coeffluk}, \refeq{coeffr}. 
  
  We finally comment on the isotropic case, $\g=0$. Here,
  $\nu_1(\t,\g=0)=\t^2/(4\pi^2)$, in agreement with the result of
  \cite{Lou03}. However, we expect that a logarithmic dependence of the
  amplitude on the distance occurs, similarly to what happens for the
  two-point functions \cite{Affleck98,Lukyanov98,Lukyanov03}. We leave the study of this case as a project for future research. 

  \subsubsection{The generalized string correlation function
    $\mathcal{O}(n,\t)$}
   
  From Eq.~\refeq{orfromrho}, the asymptotics of $\mathcal O(n,\t)$ is obtained once the asymptotics of
  $\rho(n,\t)$ is known. It is nevertheless instructive to perform a
  consistency check of this result by calculating the asymptotics of $\mathcal
  O$ directly by using field-theoretical arguments. 

  Therefore, one might be tempted to take the asymptotic
  expansion of $S^z(x)$, Eq.~\refeq{szas}, and insert it into
  Eq.~\refeq{ordef}. However, in such a calculation the leading terms given in
  Eq.~\refeq{ras} would be absent. We are thus lead to use the following
  asymptotic expansion for the $S^z$-operators at sites $1$ and $n$ involved in $\Or(n,\t\neq 0)$:
  \be
  S^z(x) &\sim &s_0+\frac{\ve}{2\pi R} \partial_x\vp(x)\nn\\
& & + \sum_{m=0}^\infty (-1)^j
  \ve^{(2m+1)^2/(4\pi R^2)}C_m \cos
  \l(\frac{2 m+1}{R} \vp(x)\r)+ \mbox{descendants }\,, \label{szas2}
  \ee
  with $x=\varepsilon j$. The asymptotic expansion starts with a finite constant $s_0$. 
  For the asymptotics of the phase factor in $\Or(n,\t)$, we still use
  Eq.~\refeq{szas}. Carrying out the same calculations as above, one finds
  $s_0^2=\tan^2(\theta/4)$, which vanishes for $\t=0$. The intriguing point is that we
  have to modify the asymptotic expansion for the spins at sites 1 and $n$ in
  $\Or(n,\t)$ without modifying the Hamiltonian. Namely, it looks as if in the
  asymptotic limit, the phase operator in $\Or(n,\t)$ acts as a local field on
  the edge spins.

  \section{Conclusion and outlook}
  We evaluated the string correlation functions $\rho(n,\t)$ and $\Or(n,\t)$
  for the critical anisotropic spin-1/2 chain. For small $n$, exact results
  were obtained from the Bethe Ansatz, whereas in the asymptotic limit, both
  the amplitudes and the exponents of the leading decay could be determined
  from field theory. The field-theoretical results agree well with the Bethe
  Ansatz data. Especially, for $\D=0$, the asymptotics could be confirmed directly 
  from the Bethe Ansatz results.

  Most interestingly, the leading decay of the two-point $xx$-correlation
  function was recovered, Eq.~\refeq{xx}. Whether this result has a physical
  background has to be clarified. As far as the limit $\t\to 0$ in $\Or(n,\t)$
  is concerned, we have recovered the expected result Eq.~\refeq{limitOr}. However, 
  the rather heuristic expansion \refeq{szas2} in the field-theory for $\theta\neq 0$ 
  deserves further investigations in the future.  
  \section*{Acknowledgments}
  We acknowledge valuable discussions with M. Batchelor, F. G\"ohmann, A. Kl\"{u}mper, M. Oshikawa, K. Sakai and
  M. Takahashi. MB is grateful for hospitality of the ISSP, University of Tokyo, where part
  of this work was carried out. Financial support from the German Research
  Council under grant number BO 2538/1-1 and from ARC Linkage International are also acknowledged (MB).

\appendix 

\section{Numerical values of string correlation functions} 
\setcounter{equation}{0}
\setcounter{table}{0}
For ${\D=1}$ and ${1/2}$, the string correlation 
functions ${\rho(n,\t)}$ and ${\Or(n,\t)}$ can be evaluated analytically up to ${n=8}$ and 
${n=9}$, respectively. Here firstly, we list their precise numerical values for 
${\theta=\pi/4,\pi/2,3\pi/4, \pi}$ up to ${n=8}$, based on these analytical expressions (see Table \ref{Delta10} and Table \ref{Delta05}). 
Note that ${\rho(1,\theta)=\cos \frac{\t}{2}}$ irrespective of ${\D}$ and therefore we have 
\be
\rho\(1,\frac{\pi}{4}\)&=& \cos \frac{\pi}{8}=  0.923880, \ \ \ \ 
\rho\(1,\frac{\pi}{2}\) = \cos \frac{\pi}{4}= 0.707107, \nonumber \\
\rho\(1,\frac{3 \pi}{4}\)&=& \cos \frac{3\pi}{8}=0.382683, \ \ \ \ 
\rho\(1,\pi\)=0. \nonumber 
\ee
Note also that ${\Or(2,\t)= -4 \bra S_1^z S_2^z \ket}$ irrespective of ${\t}$ by its definition.

\begin{table}[h]
\begin{center}
\begin{small}
\begin{tabular}
{@{\hspace{\tabcolsep}\extracolsep{\fill}}ccccccccc} 
\hline
 $n$ & 2 & 3 & 4 & 5 & 6 & 7 & 8 \\ 
\hline 
$\rho(n,\pi/4)$ 
&0.940083&0.915627&0.925111&0.910171&0.917092
&0.90616&0.911707\\
$\rho(n,\pi/2)$ 
&0.795431&0.685542&0.744898&0.671293&0.718266
&0.66085&0.70065\\
$\rho(n,3\pi/4)$ 
&0.65078&0.362761&0.565509&0.349604&0.521325&
0.339972&0.492564\\
$\rho(n,\pi)$ 
&0.590863&0       &0.491445&0       &0.440302
&0      &0.407242\\
$\Or(n,\pi/4)$ &
 0.590863 & -0.224243 & 0.24353 & -0.120692 & 0.170343 & -0.0786391 & 0.137344 \\
$\Or(n,\pi/2)$ &
 0.590863 & -0.171628 & 0.34622 & -0.0787594 & 0.282725 & -0.038569 & 0.2504 \\
$\Or(n,3\pi/4)$ &
 0.590863 & -0.0928846 & 0.44891 & -0.0352563 & 0.394313 & -0.00917944 & 0.361674 \\
\hline
\end{tabular}
\end{small}
\end{center}
\caption{Exact values of string correlation functions for ${\D=1}$}
\label{Delta10}
\end{table}

\begin{table}[h]
\begin{center}
\begin{small}
\label{diag}
\begin{tabular}
{@{\hspace{\tabcolsep}\extracolsep{\fill}}ccccccccc} 
\hline
 $n$ & 2 & 3 & 4 & 5 & 6 & 7 & 8 \\ 
\hline 
$\rho(n,\pi/4)$ &
 0.926777 & 0.909081 & 0.909299 & 0.900034 & 0.899811 & 0.893662 & 0.893337  \\
$\rho(n,\pi/2)$ &
 0.75 & 0.668437 & 0.692139 & 0.644847 & 0.661899 & 0.628299 & 0.641883  \\
$\rho(n,3\pi/4)$ &
 0.573223 & 0.346957 & 0.477542 & 0.32521 &  0.429591 & 0.310018 & 0.398987  \\
$\rho(n,\pi)$ &
  0.5 & 0. & 0.389404 & 0. & 0.335008 & 0. & 0.30088\\
$\Or(n,\pi/4)$ &
  0.5 & -0.101049 & 0.14059 & -0.0285505 & 0.0880858 & -0.00509166 & 0.0689805 \\
$\Or(n,\pi/2)$ &
  0.5 & -0.0773398 & 0.243652 & 0.000467493 & 0.192033 & 0.0293118 & 0.168568  \\
$\Or(n,3\pi/4)$ &
  0.5 & -0.041856 & 0.346715 & 0.012332 & 0.29362 & 0.033798 & 0.263161  \\
\hline
\end{tabular}
\end{small}
\end{center}
\caption{Exact values of string correlation functions for ${\D=1/2}$}
\label{Delta05}
\end{table}  
For ${\D=1/2}$ let us compare the results above with the numerical value of the asymptotic 
formulae Eq.~\eqref{ras} and Eq.~\eqref{oasym} with ${\g=\pi/3}$ in Table \ref{Delta05Asym}.
\begin{table}[h]
\begin{center}
\begin{small}
\begin{tabular}
{@{\hspace{\tabcolsep}\extracolsep{\fill}}ccccccccc} 
\hline
 $n$ & 2 & 3 & 4 & 5 & 6 & 7 & 8 \\ 
\hline 
$\rho_{\rm Asym}(n,\pi/4)$ &
0.926694 & 0.909865 & 0.909106 & 0.900388 & 0.899692 & 0.893859 & 0.893259 
\\
$\rho_{\rm Asym}(n,\pi/2)$ &
 0.751733 & 0.669839 & 0.692208 & 0.645454 & 0.661862 & 0.628622 & 0.641843  
\\
$\rho_{\rm Asym}(n,3\pi/4)$ &
 0.577912 & 0.348016 & 0.478151 & 0.325667 & 0.429729 & 0.310258 & 0.399020 
\\
$\rho_{\rm Asym}(n,\pi)$ &
 0.506119 & 0 & 0.390271 & 0 & 0.335222 & 0 & 0.300940 
\\
$\Or_{\rm Asym}(n,\pi/4)$ &
 0.306262 & -0.0812361 & 0.116753 & -0.0204905 & 0.0793274 & -0.000774240 & 0.0644446       
\\
$\Or_{\rm Asym}(n,\pi/2)$ &
 0.402260 & -0.0722680 & 0.233892 & 0.00275605 & 0.1888360 &  0.0305388 & 0.167037  
\\
$\Or_{\rm Asym}(n,3\pi/4)$ &
 0.477679 &-0.0413030 & 0.344917 & 0.0128694 & 0.293024 & 0.0341302 & 0.262857
\\
\hline
\end{tabular}
\end{small}
\end{center}
\caption{Asymptotic formulas of string correlation functions for ${\D=1/2}$}
\label{Delta05Asym}
\end{table}

We find the exact values and the asymptotic formulas are in good agreement especially for ${\rho(n,\theta)}$. 
The deviation is somewhat larger for ${\Or(n,\t)}$, for which we probably need higher order corrections to 
the asymptotic formulas to achieve better agreement. 

To confirm our asymptotic formula further, we have calculated ${\rho(n,\t)}$ numerically for several values 
of ${\D (= \pm 0.3,\pm 0.7,-0.5)}$ by means of the exact diagonalization for finite systems ${N=20 \sim 28}$.  Then we have applied an extrapolation according to $c_0+c_1/N^2+c_2/N^3+c_3/N^4+c_4/N^5$ and estimated ${\rho_{\rm Num}(n,\t)}$ in the thermodynamic limit. These values are compared with our asymptotic formula ${\rho_{\rm Asym}(n,\t)}$ in Tables \ref{Delta03}-\ref{Deltam07}. We conclude that our asymptotic formula gives fairly precise values for all ranges of ${\D}$ in the critical region.

\begin{table}[h]
\begin{center}
\begin{small}
\begin{tabular}
{@{\hspace{\tabcolsep}\extracolsep{\fill}}ccccccccc} 
\hline
 $n$ & 2 & 3 & 4 & 5 & 6 & 7 & 8 \\ 
\hline 
$\rho_{\rm Num}(n,\pi/4)$ &
 0.921405 & 0.905433 & 0.902971 & 0.894779 & 0.892836 & 0.887437 & 0.885867 
\\
$\rho_{\rm Asym}(n,\pi/4)$ &
 0.921707 & 0.905979 & 0.902974 & 0.894999 & 0.892828 & 0.887552 & 0.885852
\\
\hline
$\rho_{\rm Num}(n,\pi/2)$ &
 0.731659 & 0.658904 & 0.671538 & 0.631168 & 0.640067 & 0.612185 & 0.619215
\\
$\rho_{\rm Asym}(n,\pi/2)$ &
 0.734432 & 0.659884 & 0.672036 & 0.631562 & 0.640252 & 0.612384 & 0.619276
\\
\hline
$\rho_{\rm Num}(n,3\pi/4)$ &
 0.541914 & 0.338150 & 0.444087 & 0.312623 & 0.395710 & 0.295270 & 0.365120
\\
$\rho_{\rm Asym}(n,3\pi/4)$ &
 0.548200 & 0.338908 & 0.445270 & 0.312938 & 0.396134 & 0.295431 & 0.365265
\\
\hline
\end{tabular}
\end{small}
\end{center}
\caption{Numerical values of ${\rho(n,\theta)}$ for ${\D=0.3}$}
\label{Delta03}
\end{table}

\begin{table}[h]
\begin{center}
\begin{small}
\begin{tabular}
{@{\hspace{\tabcolsep}\extracolsep{\fill}}ccccccccc} 
\hline
 $n$ & 2 & 3 & 4 & 5 & 6 & 7 & 8 \\ 
\hline 
$\rho_{\rm Num}(n,\pi/4)$ &
 0.932056 & 0.912068 & 0.915471 & 0.904519 & 0.906536 & 0.899088 & 0.900492
\\
$\rho_{\rm Asym}(n,\pi/4)$ &
 0.931684 & 0.913012 & 0.915038 & 0.905020 & 0.906242 & 0.899399 & 0.900275
\\
\hline
$\rho_{\rm Num}(n,\pi/2)$ &
 0.768025 & 0.676242 & 0.712563 & 0.656541 & 0.683506 & 0.642401 & 0.664320
\\
$\rho_{\rm Asym}(n,\pi/2)$ &
 0.768824 & 0.677927 & 0.712076 & 0.657389 & 0.683096 & 0.642903 & 0.663983
\\
\hline
$\rho_{\rm Num}(n,3\pi/4)$ &
 0.603994 & 0.354168 & 0.511302 & 0.335990 & 0.464170 & 0.322975 & 0.433890
\\
$\rho_{\rm Asym}(n,3\pi/4)$ &
 0.607173 & 0.355419 & 0.511188 & 0.336608 & 0.463845 & 0.323332 & 0.433544
\\
\hline
\end{tabular}
\end{small}
\end{center}
\caption{Numerical values of ${\rho(n,\theta)}$ for ${\D=0.7}$}
\label{Delta07}
\end{table}

\begin{table}[h]
\begin{center}
\begin{small}
\begin{tabular}
{@{\hspace{\tabcolsep}\extracolsep{\fill}}ccccccccc} 
\hline
 $n$ & 2 & 3 & 4 & 5 & 6 & 7 & 8 \\ 
\hline 
$\rho_{\rm Num}(n,\pi/4)$ &
 0.903373 & 0.887893 & 0.878937 & 0.870856 & 0.865109 & 0.859693 & 0.855477
\\
$\rho_{\rm Asym}(n,\pi/4)$ &
 0.903927 & 0.887947 & 0.879042 & 0.870896 & 0.865149 & 0.859714 & 0.855487 
 \\
\hline
$\rho_{\rm Num}(n,\pi/2)$ &
 0.670096 & 0.61307 & 0.597811 & 0.569629 & 0.559974 & 0.541915 & 0.534903 
 \\
$\rho_{\rm Asym}(n,\pi/2)$ &
 0.674020 & 0.613330 & 0.598582 & 0.569822 & 0.560266 & 0.542024 & 0.535013
\\
\hline
$\rho_{\rm Num}(n,3\pi/4)$ &
 0.436818 & 0.295804 & 0.332446 & 0.256666 & 0.283712 & 0.232404 & 0.253962
\\
$\rho_{\rm Asym}(n,3\pi/4)$ &
 0.446622 & 0.296198 & 0.334401 & 0.256936 & 0.284437 & 0.232560 & 0.254249
\\
\hline
\end{tabular}
\end{small}
\end{center}
\caption{Numerical values of ${\rho(n,\theta)}$ for ${\D=-0.3}$}
\label{Deltam03}
\end{table}

\begin{table}[h]
\begin{center}
\begin{small}
\begin{tabular}
{@{\hspace{\tabcolsep}\extracolsep{\fill}}ccccccccc} 
\hline
 $n$ & 2 & 3 & 4 & 5 & 6 & 7 & 8 \\ 
\hline 
$\rho_{\rm Num}(n,\pi/4)$ &
 0.895942 & 0.877985 & 0.866417 & 0.857132 & 0.849944 & 0.843718 & 0.838522 
\\
$\rho_{\rm Asym}(n,\pi/4)$ &
 0.895477 & 0.877611 & 0.866290 & 0.857039 & 0.849893 & 0.843674 & 0.838481
\\
\hline
$\rho_{\rm Num}(n,\pi/2)$ &
 0.644723 & 0.587178 & 0.562522 & 0.535287 & 0.520367 & 0.503241 & 0.492689
\\
$\rho_{\rm Asym}(n,\pi/2)$ &
 0.646724 & 0.586707 & 0.562879 & 0.535292 & 0.520512 & 0.503266 & 0.492719
\\
\hline
$\rho_{\rm Num}(n,3\pi/4)$ &
 0.393503 & 0.271884 & 0.284932 & 0.226344 & 0.236472 & 0.199498 & 0.207607
\\
$\rho_{\rm Asym}(n,3\pi/4)$ &
 0.402915 & 0.271927 & 0.286782 & 0.226578 & 0.237163 & 0.199648 & 0.207867
\\
\hline
\end{tabular}
\end{small}
\end{center}
\caption{Numerical values of ${\rho(n,\theta)}$ for ${\D=-0.5}$}
\label{Deltam05}
\end{table}

\begin{table}[h]
\begin{center}
\begin{small}
\begin{tabular}
{@{\hspace{\tabcolsep}\extracolsep{\fill}}ccccccccc} 
\hline
 $n$ & 2 & 3 & 4 & 5 & 6 & 7 & 8 \\ 
\hline 
$\rho_{\rm Num}(n,\pi/4)$ &
 0.886810 & 0.863491 & 0.847596 & 0.835540 & 0.825990  
 & 0.818029 & 0.811263
\\
$\rho_{\rm Asym}(n,\pi/4)$ &
 0.883755 & 0.861533 & 0.846496 & 0.834864 & 0.825552 
 & 0.817718 & 0.811018
\\
\hline
$\rho_{\rm Num}(n,\pi/2)$ &
 0.613546 & 0.549304 & 0.512895 & 0.483364 & 0.462734 &
 0.444631 & 0.430668
\\
$\rho_{\rm Asym}(n,\pi/2)$ &
 0.610618 & 0.545990 & 0.511485 & 0.482427 & 0.462241 & 
 0.444263 & 0.430380
\\
\hline
$\rho_{\rm Num}(n,3\pi/4)$ &
 0.340281 & 0.236892 & 0.225198 & 0.182532 & 0.177268 & 
 0.153088 & 0.150087
\\
$\rho_{\rm Asym}(n,3\pi/4)$ &
 0.348075 & 0.235495 & 0.226127 & 0.182395 & 0.177655 & 
 0.153117 & 0.150189 
\\
\hline
\end{tabular}
\end{small}
\end{center}
\caption{Numerical values of ${\rho(n,\theta)}$ for ${\D=-0.7}$}
\label{Deltam07}
\end{table}

\section{Proof of \refeq{limit}}
\setcounter{equation}{0}
We prove Eq.~\refeq{limit} in the form 
\begin{align}
\lim_{\theta \to 0} D(2\pi-\theta) \frac{4 \pi^2}{\theta^2} = \left[\frac{\Gamma\left( \frac{\eta}{2-2 \eta} \right)}{2 \sqrt{\pi} 
\Gamma\left(\frac{1}{2-2\eta} \right)} \right]^{1/\eta} 
\exp\left[ \int_{0}^{\infty} \left( \frac{\sinh((2\eta-1)t)}{\sinh \eta t \cosh((1-\eta)t)} 
- \frac{2 \eta-1}{\eta} {\rm e}^{-2t} \right) \frac{{\rm d} t}{t} \right]. \label{B-1}
\end{align}
By taking the logarithm and introduce a variable $z\equiv\theta/{2\pi}$, we can calculate 
the LHS from the definition Eq.~\eqref{coeffluk} as follows. 
\begin{align}
&\lim_{\theta \to 0} \ln \left( D(2\pi-\theta) \frac{4 \pi^2}{\theta^2} \right) \nonumber \\ 
&=  \ln \left[\frac{\Gamma\left( \frac{\eta}{2-2 \eta} \right)}{2 \sqrt{\pi} 
\Gamma\left(\frac{1}{2-2\eta} \right)} \right]^{1/\eta} -\lim_{z \to 0} \left\{ \int_{0}^{\infty} \left( \frac{\sinh^2 (1-z)t}{\sinh t \cosh(1-\eta)t \sinh \eta t} - \frac{(1-z)^2}{\eta} {\rm e}^{-2t} \right) \frac{{\rm d} t}{t} 
+ 2 \ln z \right\} \label{2}
\end{align}
Now substitute the function ${\ln z}$ by its integral represention
\begin{align}
\ln z = \int_{0}^{\infty} ({\rm e}^{-t} - {\rm e}^{-z t}) \frac{{\rm d} t}{t} 
=  \int_{0}^{\infty} ({\rm e}^{-2t} - {\rm e}^{-2z t}) \frac{{\rm d} t}{t}, \ \ \ \ \ \ \ \ \ (\Re (z) >0)
\end{align}
we have 
\begin{align}
& -\lim_{z \to 0} \left\{ \int_{0}^{\infty} \left( \frac{\sinh^2 (1-z)t}{\sinh t \cosh(1-\eta)t \sinh \eta t} - \frac{(1-z)^2}{\eta} {\rm e}^{-2t} \right) \frac{{\rm d} t}{t} 
+ 2 \ln z \right\}  \nonumber  \\
=& -\lim_{z \to 0} \left\{ \int_{0}^{\infty} \left( \frac{\sinh^2 (1-z)t}{\sinh t \cosh(1-\eta)t \sinh \eta t} -2 {\rm e}^{-2zt} +\left(2 - \frac{(1-z)^2}{\eta}\right) {\rm e}^{-2t} \right) \frac{{\rm d} t}{t} \right\} \nonumber \\
=& \int_{0}^{\infty} \left( \frac{-\sinh t}{\cosh(1-\eta)t \sinh \eta t} +2  -\frac{2 \eta-1}{\eta} {\rm e}^{-2t} \right) \frac{{\rm d} t}{t} \nonumber \\
=& \int_{0}^{\infty} \left( \frac{\sinh((2\eta-1)t)}{\cosh(1-\eta)t \sinh \eta t} -\frac{2 \eta-1}{\eta} {\rm e}^{-2t} \right) \frac{{\rm d} t}{t}.
\end{align}
Thus Eq.\refeq{B-1}, namely, Eq.\refeq{limit} is proved.

\section{Proof of \refeq{coeffDelta0}}
\setcounter{equation}{0}
If we use the notation $z=\t/(2\pi)$, the asymptotic amplitude of the string correlation function (\ref{coeffluk}) is rewritten as
\begin{align}
D(\t,\g)=\[\frac{\G\(\frac{\eta}{2-2\eta}\)}{2\sqrt{\pi}\G\(\frac1{2-2\eta}\)}\]
^{z^2/\eta}
\exp\[-\ip\(\frac{\sh^2zt}{\sh t\ch(1-\eta)t\sh\eta t}
-\frac{z^2\te^{-2t}}{\eta}\)\frac{\d t}{t}\].
\end{align}
Setting the parameters as 
\begin{align}
\g=\pi/2, \quad \D=\cos\g=0, \quad \eta=(\pi-\g)/\pi=1/2,
\end{align}
we obtain
\begin{align}
D(\t,\pi/2)&=\(\frac12\)^{2z^2}\exp\[-\ip\(\frac{\sh^2(zt)}{\sh t\ch(t/2)\sh(t/2)}
-2z^2\te^{-2t}\)\frac{\d t}{t}\]\nn\\
&=4^{-z^2}\exp\[-2\ip\(\frac{\sh^2(zt)}{\sh^2t}-z^2\te^{-2t}\)\frac{\d t}t\].
\end{align}
On the other hand, the Barnes $G$-function enjoys the following integral representation \cite{Vigneras61}
\begin{align}
\ln G(1+z) =&
\ip\frac{\te^{-t}}{t(1-\te^{-t})^2}\(1-zt+\frac{z^2t^2}2-\te^{-zt}\)\d t \nonumber \\
& -(1+\g_E)\frac{z^2}2
+\(\log\frac{2\pi}\te\)\frac{z}2, \ \ \ \ \ \ \ \ (\Re(z)>-1) 
\end{align}
where $\g$ is the Euler-Mascheroni constant. From this integral representation we have
\begin{align}
-\ln \l[G(1+z)G(1-z)\r]
&=\ip\frac{\te^{-t}}{t(1-\te^{-t})^2}\(-2-z^2t^2+2\ch(zt)\)\d t+(1+\g_E)z^2\nn\\
&=\ip\frac{-2-z^2t^2+2\ch(zt)}{4\sh^2(t/2)}\frac{\d t}t+(1+\g_E)z^2\nn\\
&=\ip\frac{-1-2z^2t^2+\ch(2zt)}{2\sh^2t}\frac{\d t}t+(1+\g_E)z^2\nn\\
&=\ip\frac{\sh^2(zt)-z^2t^2}{\sh^2t}\frac{\d t}t+(1+\g_E)z^2.
\label{coeffIntegral}
\end{align}
Now substituting the formula
\begin{align}
\ip \left( \frac{t}{\sinh^2 t} - \frac{\te^{-t}}{\sinh t} \right)\d t
&= \left[ \l(\ln (\sinh t) - t \coth t \right) 
- \left( \ln \left( \sinh t \right) -t \right) \right]_{0}^{\infty} \nn \\ 
&= \left[ \frac{-2 t}{\te^{2 t}-1} \right]_{0}^{\infty} =1,
\end{align}
into the standard integral representation of the Euler-Mascheroni constant,
\begin{align}
\g_E=\ip  \left(\frac{\te^{-t}}{1-\te^{-t}} - \frac{\te^{-t}}{t}\right){\rm d} t
\end{align}
we can establish another integral formula
\begin{align}
1+\g_E=\ip\(\frac{t^2}{\sh^2t}-\te^{-2t}\)\frac{\d t}t. \label{newEulerGammaIntegral}
\end{align}
Substituting Eq.~\refeq{newEulerGammaIntegral} into Eq.~\refeq{coeffIntegral} yields
\begin{align}
-\ln\l[G(1+z)G(1-z)\r]=\ip\(\frac{\sh^2(zt)}{\sh^2t}-z^2\te^{-2t}\)\frac{\d t}t,
\end{align}
from which we conclude
\begin{align}
D(\t,\pi/2)=4^{-z^2}\[G(1+z)G(1-z)\]^2,
\end{align}
which is equivalent to Eq.~\refeq{coeffDelta0}.

\end{document}